\documentclass{iopart}
\usepackage{amssymb}
\usepackage{amsmath}
\usepackage{graphicx}
\usepackage{url}
\usepackage{setspace}
\usepackage{subfig}

\newcommand{\includefigure}[1][!]{\def\figXDim{#1}\figSubCmd}
\newcommand{\figSubCmd}[2][!]{\resizebox{\figXDim}{#1}{\includegraphics[type=pdf
,ext=.PDF,read=.PDF]{#2}}}

\newlength{\myfigwidth}
\begin{document}
\title[The Critical Coupling Likelihood Method]{The Critical Coupling
  Likelihood Method: A new approach for
  seamless integration of environmental and operating conditions of
  gravitational wave detectors into gravitational wave searches.} 
\author{Cesar A. Costa$~^1$ and Cristina V. Torres$~^2$}
\address{
$~^1$Louisiana State University \\
Department of Physics \\
202 Nicholson Hall, Tower Dr. \\
Baton Rouge, LA 70803-4001 USA \\
$~^2$LIGO Livingston Observatory \\
PO Box 940 \\
Livingston, LA 70754 USA}
\ead{\mailto{cesar.costa@ligo.org}\\\mailto{cristina.torres@ligo.org}}

\begin{abstract}
Any search effort for gravitational waves (GW) using interferometric
detectors like LIGO needs to be able to identify if 
and when noise is coupling into the detector's output signal.
The Critical Coupling Likelihood (CCL) method has been developed to
characterize potential noise coupling and in the
future aid GW search efforts. 
By testing two hypotheses about pairs of channels, CCL is able to
identify undesirable coupled instrumental noise from potential
GW candidates.  Our preliminary results show that CCL can associate up
to $\sim 80\%$ of observed artifacts with $SNR \geq 8$, to local noise
sources, while reducing the duty cycle of the instrument by $\lesssim
15\%$.  An approach like CCL will become increasingly important as GW
research moves into the Advanced LIGO era, going from the first GW
detection to GW astronomy.
\end{abstract} 

\maketitle

\section{Introduction}
Detectable gravitational waves (GWs) are perturbations of the
local space-time metric which are associated with distant astrophysical
phenomena. According to general relativity, these perturbations travel at the
speed of light and are generated by astronomical scale masses with time varying
quadrupolar and higher moments. The main goal of the Laser Interferometric
Gravitational-Wave Observatory (LIGO) is to detect GWs and use this
information to study the astrophysics associated with those sources
\cite{LIGOurl}.

The LIGO detectors are now undergoing a major 
upgrade. This upgrade will increase the instrument's sensitivity,
extending the binary neutron star (BNS) observational range from the current
$\sim20\rm{Mpc}$ to $200 \rm{Mpc}$ \cite{aLIGO2}. This change in the
observational volume should increase the expected observable rate of
GW signals.  Using an improved network of GW detectors, one can expect
BNS detection rates to improve from $1/50\,\rm{yr}^{-1}$ to $40\,\rm{yr}^{-1}$ 
\cite{EventRates1,EventRates2}.  In order to achieve such sensitivity,
LIGO instruments will be completely refitted with advanced components
and control systems. This new configuration is called Advanced LIGO
(aLIGO).  During the aLIGO operational era, the first GW detection
will mark the beginning of gravitational wave astronomy.  

The first GW detection will be thoroughly reviewed.
In an era of regular GW detection, human resource costly 
procedures such as the traditional follow-ups and post-facto data quality
(DQ) studies should be relied on infrequently; the state of the
instrument, and hence the DQ have a significant impact on the
effectiveness of LIGO's GW searches.  Current DQ efforts identify
epochs of questionable data quality.  The DQ is analyzed in subsets,
starting with the best data available, and epochs of decreasing
quality data are added to the analysis to increase the effective
observing time of the GW detector \cite{HowToDQ,ChannelSafety02}.
Upon completion or near the end of an analysis the results are either 
disregarded immediately or further scrutinized during the follow-up
procedure, to determine if the potential GW candidate is the signature
of an actual GW signal \cite{FollowupWorkshop,flwupGouaty}.

In this paper we are proposing a new method, the Critical Coupling
Likelihood method, to investigate instrumental behavior and DQ.  This method
is intended to quantify instrumental operating conditions. We will
also remark on how this information could be integrated into future GW
searches.  The CCL method uses the same sources of information as
current DQ investigation methods, though it is a significant
departure from those methods \cite{DataQuality}.  LIGO's
current DQ methods have been invaluable and used in some form for all
previous science runs for studying instrumental behavior and noise
coupling. These existing methods became the standard for
investigating LIGO's DQ but in an aLIGO era they may no longer be 
optimal methods for future DQ studies.  Significant revisions to these
methods or new methods may be needed in order to make DQ investigations
manageable in an aLIGO era \cite{2012LscWhitePaper}.  

Understanding preponderant non-Gaussian noise sources is compulsory so
that one can isolate them during the search process without
unnecessarily compromising the instrument duty cycle.  These noise sources
generate transients which are typically the result of a local influence on the
instrument, and can lead to false GW candidates.  In order to uncover the
sources of these transients in LIGO data there are a large number of
sensors which are collectively referred to as physical and environmental
monitors (PEMs). 

LIGO records information about the detector's operating 
environment and the control systems of the interferometer.  This
information is used to determine if a specific sampled time from the
instrument's GW output signal is or is not a noise transient in the system.
In order to accomplish this, each of the sensors are analyzed individually to
identify departures from their nominal behavior.  The results of this analysis
could be used to identify and discard noise transients occurring during times of
questionable data quality as part of a GW search
\cite{S1flowchart}.

We propose to improve GW searches by applying our
method, the Critical Coupling Likelihood (CCL) method.
This method is expected to integrate information about the
instrument's operating condition and its environment directly into a
search.

\section{The Critical Coupling Likelihood}
CCL is a statistical method intended to quantitatively identify as
many avenues of environmental-to-instrumental coupling as possible.
To accomplish this, the CCL method is based on time coincidence between event
pairs from preselected data streams.  This pairing is done between the GW data
stream (GW channel) and an auxiliary sensor data stream (sensor
channel).  This method is intended to distinguish
between real coincidences (coupling) and accidental coincidences,
which are coincidences unlikely to be of physical interest.

It is reasonable to expect that any sensor will have some level of
inherent noise.  It is important that this noise be Gaussian in
nature, and for well engineered equipment this is typically the case.
CCL has been designed to be as insensitive as possible to the inherent Gaussian
noise properties of a pair of sensors.  This insensitivity to
the unrelated Gaussian behavior between the data sets 
makes it possible to distinguish accidental coincidence from suspected
coupling.

GW and sensor channels are analyzed pairwise to identify
interesting potential artifacts; the specific algorithm used is a matter of
convenience.  An artifact for the purposes of a CCL analysis is an
intermediate data product resulting from the processing of raw input
data, like a time series analyzed using a time frequency (TF)
decomposition.  In this case the artifact is a statistically
interesting structure identified in the time frequency plot of raw
input data.  The artifacts used have properties like time of
occurrence, central frequency, bandwidth, estimated signal-to-noise
ratio (SNR), etc. 

CCL uses at least two artifact properties, and for the purposes of this paper we
chose to use the SNR $\rho$ and time of occurrence $t$.  One could use other
artifact properties and by using any combination of these measurements
one can construct varying models. All of these variants in principle
can be used to identify suspected coupling. 

Differentiating unrelated artifacts (accidental coincidences)
from coupled ones relies on the creation of two models. The
potentially interesting set of artifacts, the coupled model
(foreground) $P_f$, describes the temporal relationship between the GW
data output and activity in the instrument's environment.  It also
contains unrelated time coincidences (accidental coincidences) for
this reason one needs a second set of artifacts. This second set, the 
uncoupled model (background) $P_b$, has no statistically meaningful
relationships between artifacts in the GW and sensor data streams.       
As the name implies CCL is a method capable of determining the
likelihood of instrumental coupling being absent or present.  In order
to make this determination one would consider the ratio of the
foreground model to a background model as follows
\begin{equation}
\label{eq:roughCCL}
CCL \propto 2\log_{10}\left(\frac{P_f}{P_b}\right).
\end{equation}
In calculating this quantity one is evaluating the potential for
coupling at a specific time, between the GW channel and an individual
sensor channel paired to it.  The two quantities $P_f$ and $P_b$ must
be constructed from the artifacts sampled from the
GW channel and an associated sensor channel respectively.

For the GW channel,  selection criteria
are applied to restrict the number of identified artifacts $(y_i)$
(i.e. frequency interval, SNR range, etc.), and this set is defined as 
\begin{equation}
\mathbf{Y} = [y_1,y_2,\ldots,y_m].
\label{eq:allArtifactY}
\end{equation}
For the sensor channel no restrictions are imposed all the
artifacts $(x_i)$ are used and this set is defined as  
\begin{equation}
\mathbf{X} = [x_1,x_2,\ldots,x_n].
\end{equation}
All data analyzed will need to be organized into sets derived from
analyzing intervals of time where the GW data is uniform in behavior.
These intervals of uniform behavior are dependent on detector
configuration changes, the operating condition of the detector, and 
changes in the underlying stationary behavior of the detector.

\section{Sampling Method}
\label{sec:samplingMethod}
Models from sufficiently sampled data sets $\mathbf{X}$ and
$\mathbf{Y}$ will be used to describe potential
relationships between the instrument and 
environmental effects.  The epoch of sampling validity is defined as a
time interval when only trivial changes to the running state of the GW
detector have occurred.  In defining epochs this way, the unknown
coupling function (system function) should be nearly stationary
between the GW detector and environment \cite{sigBook}.   The minimum
amount of data required to build a reliable model is directly
proportional to the rate of the sensor artifacts recorded and as such, 
the total amount of aggregated time (samples) required to build each
model will be sensor dependent.    

We can choose to use a small subset of artifacts from all 
available detector and associated sensor data.  The potential volume
of available artifacts to use in this analysis is prohibitively large,
and we want to restrict the size of the data sets to be only
representative of bad detector behavior.  This restriction can be
achieved by imposing a uniform sampling in time of the detector and
sensor data streams instead of processing all the available data.
Uniform sampling involves placing a minimum threshold on artifact SNR
and the rate at which we collect artifacts from the data.
Ideally the best course of action is to preferentially sample time
intervals the detector and sensor data appear to be noisier than
normal.   It is nontrivial to identify noisier than usual intervals of
time, but by using a targeted sampling approach one can create models
that contain data from periods where the instrument appears to more
noisy than usual. Periods of more noisy than usual data can be
characterized by a noticeable increase in the rate of artifacts
identified per unit time our data sets.  By proactively selecting
artifacts during these noisier periods a smaller number of sampled
artifacts can be used to identify coupled noise sources.

\section{Model Resolution}
\label{sec:ModelResolution}
The statistical properties of the collected data sets can be
expressed by a probability mass function (PMF)---the discrete form of a
probability density function (PDF).  Each model (coupled and
uncoupled) is represented  by a two dimensional (2D) conditional probability 
distribution (CPD) which associates SNRs in the GW channel and the
sensor channel.  The resolution of the CPD, is
dependent on the observation epoch and the sensor channel paired with the GW
channel.  Due to the discrete nature of the data, the resolution of
the CPDs and PMFs are finite. 

The artifacts contained in data from either the GW or sensor, can be
described by an unknown number of distributions.  One can assume that
real instruments should have a baseline noise component plus 
possibly one or more additional unknown distributions.  The baseline
noise should be Gaussian distributed, and after processing it with a
TF decomposition algorithm, the associated artifacts will follow a Rayleigh
distribution \cite{Rayleigh1}.  This distribution describes the largest
component of the observed noise. Due to the large number of low SNR
artifacts a minimum SNR threshold $\rho_0$, is applied to keep the volume
of recorded artifacts down, and anything that would be below this value
is assumed to be part of the inherent instrumental noise.  

The following equation expresses the censored form of the Rayleigh distribution
of SNRs as 
\begin{equation}
R(x) = 
\begin{cases}
\frac{x}{\sigma^2}e^{-\tfrac{x^2}{2\sigma^2}} &,\,
x>\rho_0 \\
 0 &,\, x\leq \rho_0 .
 \end{cases}
\label{eq:simpleRayleigh}
\end{equation}
where $\sigma$ is the shape parameter for the Rayleigh distributed
SNRs of Gaussian noise where the shape parameter is typically of the
order of one. The unavoidable censoring, a loss of function
support for SNRs less than $x_0$, alters the way this distribution is
normalized and will be discussed later in the paper.
 
\begin{figure}[htp]
\centering
\includefigure[0.75\textwidth]{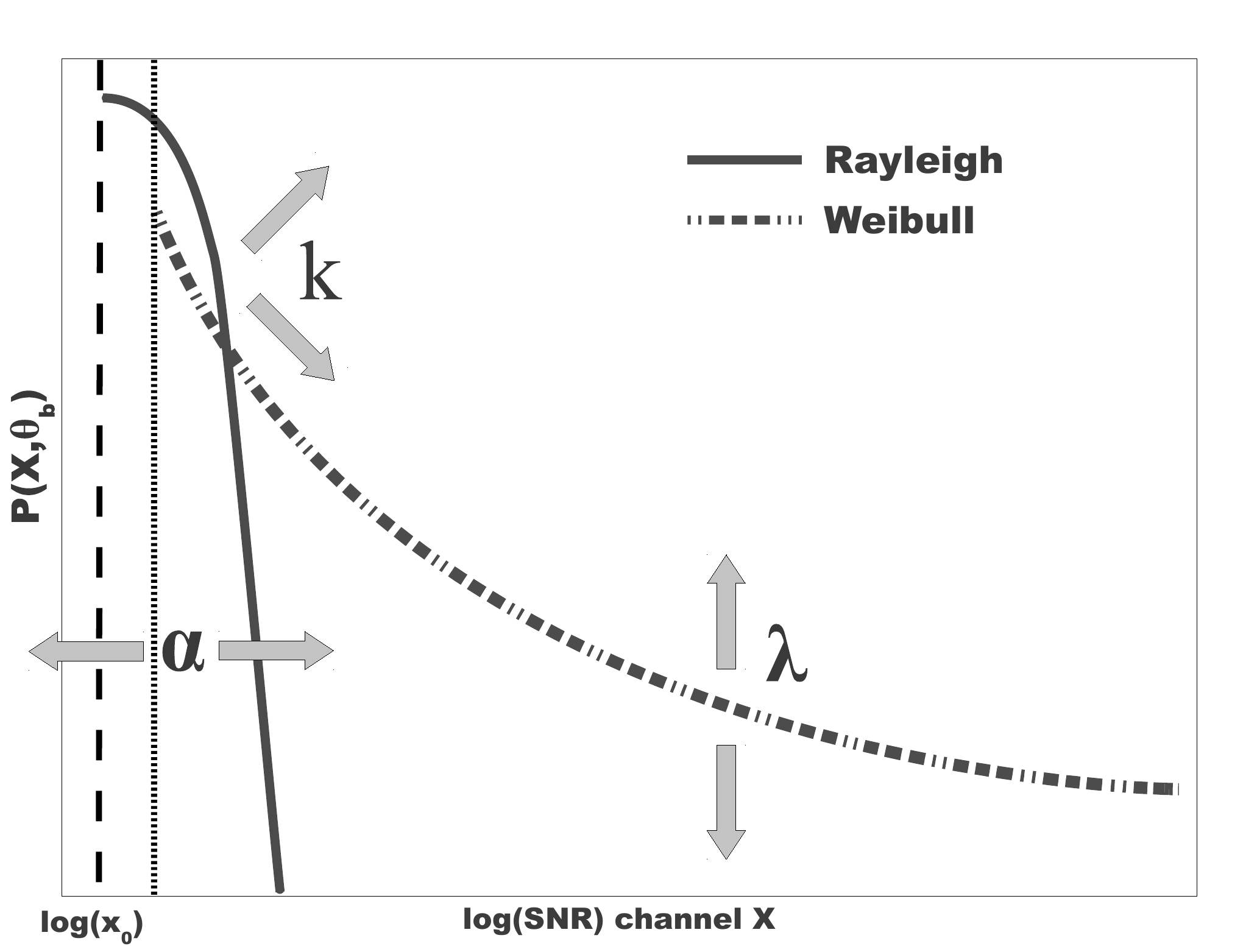}
\caption{Background Model. We have determined empirically that
  background is composed of at least two distributions.  The
  distributions shown start with a Rayleigh and is followed by a
  second distribution.  The second distribution is assumed to be a
  Modified Weibull distribution.  Together these model the
  distribution of CCL artifacts.}
\label{fig:idealModel}
\end{figure}

In addition to the Gaussian noise component observed in LIGO data
there is also an excess of outlier artifacts.  It is reasonable to 
describe this excess as at least one additional distribution \footnote{This
second distribution is observed in LIGO data. It can not be explained by
Gaussian processes. It is mostly due to instrumental malfunction and
environmental causes.}. In
reviewing a variety of data sets we empirically
determined that the most reasonable way to describe the
observed high SNR outliers is a modified Weibull
distribution (MWD) \cite{LifeDataAnalysis}.  The Weibull distribution
used for our purposes varies slightly from the standard form for a MWD
and it is described as follows: 
\begin{equation}
W(z(x)|\alpha)=
\begin{cases}
 \frac{k}{\lambda} \left( \frac{z(x)-\log{\alpha}}{\lambda} \right)^{k-1}
e^{-\left(\tfrac{z(x)-\log{\alpha}}{\lambda}\right)^k} &,\, z(x)-\alpha>0 \\
 0 &,\, z(x)-\alpha\leq 0  
 \end{cases}
\label{eq:simpleWeibull} 
\end{equation}
where $z(x)=log(x)$.
The profile of such a distribution is
determined by three parameters: the shape parameter $k$ where $k>0$;
the scale parameter $\lambda$ where $\lambda>0$; the shift
parameter $\alpha$ where $\alpha \geq 0$.  The parameter $\alpha$
scales the overlap of distributions.  A properly selected
$\alpha$ parameter allows us to adjust the relative positions of the
expected Rayleigh and Weibull distributions for an effective single
probability distribution.  Figure \ref{fig:idealModel} shows a cartoon
of how the these distributions might look relative to each other.  In
some extreme cases, the dual distribution assumption will break down
and other distributions can be added to cope with this breakdown.  The
method described in this paper is easily generalized to integrate
additional $W_{i}(z(x)|\alpha_{i})$ distributions to combat this
occasional method breakdown, but this generalization can become
computationally  challenging \cite{MultipleDistroFit}.

We defined this custom probability distribution function 
$C(x|\alpha)$, as 
\begin{equation}
C(x|\alpha) = \psi_1 R(x) + \psi_2 W(z(x)|\alpha),
\label{eq:cclPDF}
\end{equation}
where $\psi_{i}$ with $i=1,2$ are scaling parameters denoting the amount
of relative Rayleigh and Weibull based artifacts.  The $\psi$ values
are expected to behave in such a way to satisfy  
\begin{equation}
\int_{\rho_0}^{x_{\rm{max}}} C(x|\alpha) = 1
\end{equation}
and the normalization of the censored $R(x)$ will be absorbed into the parameter
$\psi_1$. 

It is necessary to compute the discrete representation of
$C(x|\alpha)$ in such a way as to best resolve all structure present.
The histogram should have the most number of bins possible, without
having an excessive number of near zero bin components.  To construct
such a histogram one needs to maximize the information entropy of the
histogram \cite{HistogramEntropyMax}.  The data sets contain a large
range of observed SNRs, as  
such, it is better to discretize this data using logarithmically
spaced bins, rather than linearly spaced ones. Consider a set of bin
edges which are base $B$ logarithmically separated, the optimal choice
of $B$ can be determined by maximizing the total histogram
entropy. Selecting $B$ can be done iteratively, 
so that the resulting histogram of the data can easily be fit to the
Rayleigh portion (low SNR) and also to the Weibull portion (high SNR)
of $C(x|\alpha)$.   To properly fit this function the entropy of the
histogram as a function of $B$ is computed. The value of $B$ is
expected to be somewhere in the interval $[2 \ldots 10)$. One 
must also decide the number of bins $b$ that should be used for
histogram construction.  In order to determine the best binning to use
one would simultaneously maximize the entropy of the histogram across
both $B$ and $b$. To discretely compute the entropy $H(x,b,B)$,
requires computing the following quantity,
\begin{equation}
H(x,b,B) = - \Sigma_{i=1}^{b} v_i(x) \ln v_i(x) \Delta x_{i}
\label{eq:findEntropy}
\end{equation}
with the following normalization 
\begin{equation}
\Sigma_{i=1}^{b} v_i(x) \Delta x_{i} = 1.
\end{equation}
The width scale of $\Delta x_{i}$ is logarithmically increasing by $B$,
and the value of $v_i(x)$ is just the normalized element count in bin
$i$.  Optimizing the logarithmic scaling to the highest bin resolution 
possible is simply a matter of identifying the global maximum of
$H(x,b,B)$.  The most accurate fit to the data is possible when using
a histogram representation with the highest possible entropy.

It is important that sampled data set properly represent the
underlying distributions, in order to create a proper fit for $C(x|\alpha)$.
Properly fitting the $C(x|\alpha)$ PDF involves  
determining the parameters, $\sigma$, $\alpha$, $\lambda$, $k$,
$\psi_1$,$\psi_2$ while respecting the threshold $\rho_0$.  The choice
of approach to determine the parameters is a
direct consequence of the data set size, and ability to properly
histogram the sampled data.  Under typical circumstances, the approach
of choice would be Quantile Maximum Product of Spacing (QMPS).  In
some cases the data sets are sparse.  For this situation we
use a fitting approach named Maximum Product of Spacing
(MPS).  The MPS approach uses all available data from the set and is
inherently more tedious to apply than the QMPS approach.  The two
different approaches converge to the same parameters for data streams
that are sufficiently sampled \cite{FitWeibull01,FitWeibull02}.

The QMPS presented itself to be the most effective approach to determine the
parameters of $C(x|\alpha)$.  The cumulative distribution function (CDF)
$c(\sigma,\alpha,k,\psi_1,\psi_2|x)$ of $C(x|\alpha)$ is
defined from the CDF of $R(x)$ 
\begin{equation}
r(\sigma|x) =
\begin{cases}
\left( 1 - e^{-\left(\tfrac{x^2}{2\sigma^2}\right)} \right)&,\,
x>\rho_0 \\
0 &,\, x\leq \rho_0
\end{cases}
\end{equation}
and from the CDF of $W(x|\alpha)$
\begin{equation}
w(\alpha,\lambda,k|z(x)) = 
\begin{cases}
1 - e^{-\left( \tfrac{z(x)-\log{\alpha}}{\lambda}\right)^k} &,\,
z(x)-\log{\alpha}>0 \\
0 &,\, z(x)-\log{\alpha}\leq 0.
\end{cases}
\end{equation}
These two equations can be combined into the following form 
\begin{equation}
c(\sigma,\alpha,k,\lambda,\rho_0,\psi_1,\psi_2|x) = 
\psi_1 + \psi_2 
- \psi_1 e^{-\left(\tfrac{x^2}{2\sigma^2}\right)} 
- \psi_2 e^{-\left( \tfrac{z(x)-\log{\alpha}}{\lambda}\right)^k}.
\label{eq:completeCDF}
\end{equation}
The parameters of interest are found by maximizing the quantity
$S(\sigma,\lambda, k,\alpha,\psi_1,\psi_2|\textbf{X})$ using QMPS as follows,
\begin{eqnarray}
S(\sigma,\lambda, k,\alpha,\psi_1,\psi_2|\textbf{X}) &=& 
\displaystyle\prod_{i=0}^{b} 
\left (
\psi_1 \left (
-e^{\left(\frac{-\hat{x}^2_i}{2\sigma^2}\right)}
+
e^{\left(\frac{-\hat{x}^2_{i-1}}{2\sigma^2}\right)}
\right )\right.\nonumber \\
&&+
\left.\psi_2 \left (
-e^{-\left(\tfrac{z(\hat{x}_i)-\log{\alpha}}{\lambda}\right)^k} +
\right.\right.\nonumber\\
&&\left.\left.e^{-\left(\tfrac{z(\hat{x}_{i-1})-\log{\alpha}}{\lambda}\right)^k}
\right )
\right )^{n_i},
\label{eq:fullQMPS}
\end{eqnarray}
and all parameters can be fit simultaneously.  Where $n_i$ is the element
count for the $i\textrm{th}$ bin. The parameter $\hat{x}_i$ represents the
$i\textrm{th}$ left side bin value which comes from the maximum
entropy histogram of data which should represent the distribution
$C(x|\alpha)$.  Using this method will give us all 
the fitting parameters required to characterize the sampled data.  

In the case that data samples are sparse, MPS is used to determine the
parameters of interest.  The MPS also seeks to maximize
the quantity $S(\sigma,\lambda, k,\alpha,\psi_1,\psi_2|\textbf{X})$,
but this approach varies as follows
\begin{eqnarray}
S(\sigma,\lambda, k,\alpha,\psi_1,\psi_2|\textbf{X}) &=& 
\displaystyle\prod_{i=0}^{j} 
\left (
\psi_1 \left (
-e^{\left(\frac{-x^2_i}{2\sigma^2}\right)}
+
e^{\left(\frac{-x^2_{i-1}}{2\sigma^2}\right)}
\right )\right.\nonumber\\
&&+
\left.\psi_2 \left (
-e^{-\left(\tfrac{x_i-\log{\alpha}}{\lambda}\right)^k} +
\right.\right.\nonumber\\
&&\left.\left.e^{-\left(\tfrac{x_{i-1}-\log{\alpha}}{\lambda}\right)^k}
\right )
\right ),
\label{eq:fullMPS}
\end{eqnarray}
where $j$ is the total number of artifacts $x_i$ used to 
fit the parameters simultaneously.  One should use QMPS whenever
possible since implementing this calculation for very large $j$
becomes computationally costly.

\section{Creating Coupled and Uncoupled Models}
Artifacts used to construct the coupled model are selected by
applying a time coincidence check, with a window of $t_{w}$, which
represents the largest absolute time difference between a GW channel
artifact and sensor channel artifact.  For this preliminary study we
chose $t_{w} = 1\;\textrm{s}$, after comparing a large
collection of sensor data artifacts from many channels with GW data
artifacts.  When analyzing large sets of sensor 
channels, it is better to choose a $t_{w}$ value which is long enough
to capture all artifacts resulting from control loop delays, and
physical delays between an environmental effect and the expected GW
detector response to this effect.

The coupled model data $P_f$ can be expressed as a
conditional probability distribution (CPD) with the following form
\begin{equation}
P_f = P(\mathbf{Y}|\mathbf{X},\theta_{\rm{f}})=\frac{P(\mathbf{Y}\bigcap
\mathbf{X},
\theta_{\rm{f}})}{P(\mathbf{X},\theta_{\rm{f}})}
\label{eq:coupledCDP}
\end{equation} 
where $P(\mathbf{Y}\bigcap\mathbf{X},\theta_{\rm{f}})$ is a two
dimensional joint probability distribution (JPD).  This JPD represents
the probability of a element in the set $Y$ (GW data), given set $X$
(sensor data) for a specific model configuration $\theta$.  In this
expression $\theta_{\rm{f}}$ (foreground) denotes the set of
coincident samples drawn during potential \textrm{coupling} times.

The data used to construct the uncoupled model is derived from all
available $\mathbf{X}$ and $\mathbf{Y}$ artifacts.  Unlike the coupled
model, the uncoupled model is intentionally designed to break the
statistical relationship between the artifacts in $\mathbf{X}$ and
$\mathbf{Y}$. Using the PDFs of $\mathbf{X}$ and $\mathbf{Y}$ a
background model JPD is built where we have forced statistical
independence.  This implies that the correct form for an uncoupled model's
JPD, $P_b$ is simply 
\begin{equation}
P_b = P(\mathbf{Y}\bigcap
\mathbf{X},\theta_{\rm{b}})=P(\mathbf{X},\theta_{\rm{b}})
\cdot
P(\mathbf{Y},\theta_{\rm{b}}).
\label{eq:simpleBackground}
\end{equation}
where $\theta_{\rm{b}}$ represents background samples used which are not
members of the foreground set identified by $\theta_{\rm{f}}$
the coupled model's CPD, the discrete forms of
\begin{math}
P(\mathbf{X},\theta_{\rm{b}})
\end{math}
and 
\begin{math}
P(\mathbf{Y},\theta_{\rm{b}})
\end{math}
must be consistent with their counterparts contained in 
\begin{math}
P(\mathbf{Y}\bigcap
\mathbf{X},\theta_{\rm{f}})
\end{math}.  In addition to knowing these functions one 
must also expect that the data which represents the coupling in the
coupled model are not the dominate data source for constructing
\begin{math}
P(\mathbf{X},\theta_{\rm{b}})
\end{math} 
and 
\begin{math}
P(\mathbf{Y},\theta_{\rm{b}}).
\end{math}   
This additional constraint is easily
satisfied when building models between any one particular channel and the GW
channel.  

\section{Calculation of CCL Function}
\label{sec:calculationCCL}
The CCL function, introduced in equation 
\ref{eq:roughCCL}, is the log likelihood ratio between the
coupled and uncoupled models previously described, and should be
expressed as follows:  
\begin{equation}
CCL(\mathbf{Y},\mathbf{X}|\theta_{\rm{f}},\theta_{\rm{b}}) = 
2\log_{10} 
\left(\frac{P(\mathbf{Y}\bigcap \mathbf{X},
\theta_{\rm{f}})}{P(\mathbf{Y}\bigcap
\mathbf{X},\theta_{\rm{b}})} 
\frac{P(\mathbf{X},\theta_{\rm{b}})}{P(\mathbf{X},\theta_{\rm{f}})}\right) .
\label{eq:cclRatioCDP}
\end{equation}
The above expression is easy to simplify algebraically because of
equation (\ref{eq:simpleBackground}) yielding 
\begin{equation}
CCL(\mathbf{Y},\mathbf{X}|\theta_{\rm{f}},\theta_{\rm{b}}) = 
2\log_{10} 
\left(
\frac{P(\mathbf{Y}\bigcap \mathbf{X},\theta_{\rm{f}})}
{P(\mathbf{Y},\theta_{\rm{b}})}
\frac{1}
{P(\mathbf{X},\theta_{\rm{f}})}
\right) .
\label{eq:cclRatioCDPSimple}
\end{equation}

The CCL function quantifies the level of suspected coupling
between a specific sensor and the GW data.  The observed CCL value can
be translated into one of three statements: no coupling suspected;
coupling suspected; or an indeterminate state with 
insufficient information to determine coupling.   In the case of no
coupling the CCL values will be negative indicating improbable
coupling.  For cases of coupling the CCL values will 
be positive indicating a potential GW channel candidate is
likely due to a localized noise source influencing the GW detector.
When the values are close to zero, either  
positive or negative, the test is incapable of determining whether the
GW candidate is a noise artifact.  

\subsection{Visualizing and Interpreting CCL Functions}
\label{sec:visualizingModels}
\begin{figure}[htp]
\begin{center}
\includefigure[0.75\textwidth]{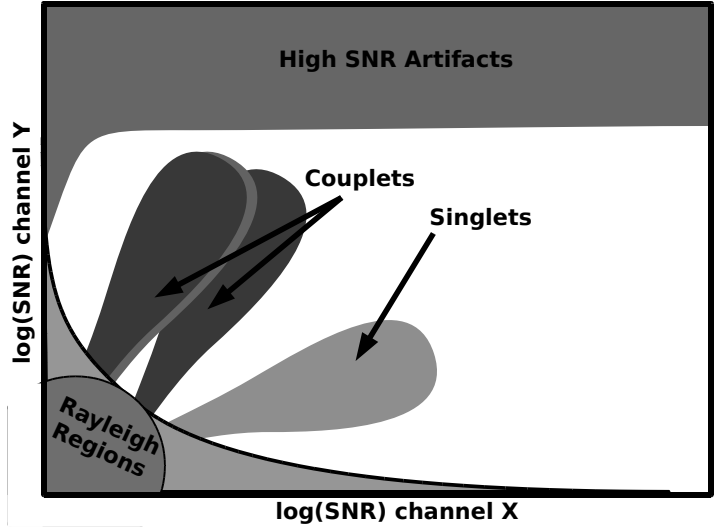}
\caption{Coupling Regions: This figure contains several shaded
  regions. These regions illustrate typical structures but not all of
  which will manifest themselves in different CCL functions. The
  Rayleigh Region, which is triangular in shape shown on the lower left
  is a region where differentiating a GW artifact from a noise
  artifact is not typically possible.  The Singlets, and Couplets
  regions indicate potential coupling which is quasi-log-linear or
  quasi-bi-log-linear in nature.  The High SNR  Artifacts region is a
  boundary region where the linearity of the observed detector
  response for those artifacts is questionable.}  
 \label{fig:cclExplainPlot}
\end{center}
 \end{figure}

Visualizing a CCL function offers insight into the characteristics of
the data.  One can consider these functions as a two dimensional color
map with the GW data, $P(\mathbf{Y})$, plotted along the vertical axis and 
the sensor data, $P(\mathbf{X})$, plotted along the horizontal
axis. When we plotted the preliminary functions we encountered distinct
patterns in the CCL values as a function of SNR 
between the two data sets.  One might expect there to be some sort of
pattern visible in plots of CCL functions and these patterns tended to be
distinct.  The distinctness of these patterns is attributable to the
coupling mechanism responsible.  Upon reviewing many CCL plots,
we noticed that they could be organized and characterized by
specifically shaped regions of high significance artifacts.   Figure
\ref{fig:cclExplainPlot} presents an illustrative cartoon
superimposing  several key regions observed from different sensor data
onto one graph.  

The color map, represents the CCL values calculated for
the data set.  All CCL functions will contain some part of or all of a 
triangular region we call \textit{Rayleigh
  Region}, composed of three joined regions, one circular and two
others triangular in shape (figure \ref{fig:cclExplainPlot}).  The
artifacts appearing in this region must be assumed to be related to
Gaussian fluctuations in either the sensor or the GW channels.  There
is no way to distinguish  Rayleigh artifacts, which are not coupled,
from potentially coupled  artifacts, which lay in one of the two
smaller triangular regions.  Artifacts in these two smaller regions
are the product of convolutions of inherent Rayleigh noise with
potentially interesting Weibull noise components.  This convolution results
in having only one 
of two pieces of useful information needed to identify coupling.  For
these two triangular regions CCL values here are typically near zero.
This is different from the shaded circular region in figure
\ref{fig:cclExplainPlot} whose artifacts are composed from two pieces
of inherent noise which is Rayleigh distributed.  For real data shown
later, this shaded circular region can be less pronounced but is
located in approximately the same region of the plot.  In this case
CCL values may take on a larger range of negative
values. The three Rayleigh related regions are inherent to all CCL
functions and typically present in some form.

The next region of interest is the \textit{Singlet} region. Here the
CCL values show what appears to be a quasi-linear relationship in a 
log-log space of observed artifact's SNR in the sensor data 
compared with the artifacts in the GW data.  Another region of
interest is \textit{Couplets} and this structure is normally seen in
pairs of CCL functions.  Consider two sensor channels that measure
similar but not identical physical phenomena like ground motion, which
can affect accelerometers and seismometers. Excessive ground motion
at the detector might produce similarly distributed artifacts in both
channels and the corresponding CCL function plots can share similar
structures which appear like adjacent or overlapping  singlets.   The
last region observed in most CCL functions, which is not sharply
defined, is the \textit{High SNR Artifacts} region.  The events that
compose this  region are at the edge of linear behavior for
differential arm motion sensing.  Because of the potential for a
non-linear instrument response, this region presents itself as
unreliable \cite{calibration}.  These plots are useful as
supplementary tools to visualize and help understand the
relationships between the sensors and the detector output.

\subsection{Disregarding CCL functions sensitive to gravitational waves}
\label{sec:disregardCCL}
The detector should in principle have its output, the GW  data, as the
only data stream which is sensitive to GWs.  In practice this 
is not guaranteed to be the case.  The detector may have sensors which
react to the presence of a GW.  It is unwise and not safe to use a
CCL function derived from sensors which may be responding to passing GWs.
One needs a prescription for identifying sensor channels that are
unsafe so those functions will not be used to identify non-GW
artifacts in the detector output \cite{ChannelSafety01}.

An unsafe CCL function, with possible undesirable sensitivity to GW
phenomena, can be identified with a simple prescription.  If the CCL
function is unsafe, then the distribution of
hardware injections identified by an individual CCL function will be similar
to the distribution of all artifacts due to hardware injections used to test
the detector.  Hardware injections are control signals introduced in the
detector which simulate its physical response to GW signals.  It is the
resulting response what is recorded in the GW channel \cite{hwInjections}. 
Using these injections it should be straightforward to identify unsafe CCL
functions.
This identification process is accomplished by using a Kolmogorov-Smirnov (KS)
test.
The null hypothesis would be, ``For a given sensor $S$,  
is the distribution of the hardware injection artifacts, identified with
$(\textrm{CCL} \geq 1)$ the same as the distribution of all known
hardware injection artifacts?''.  A CCL function is tested for safety
using 
\begin{equation} 
\sqrt{\frac{mn}{m+n}}\underset{x}{\textrm{sup}}
\left | 
F_{\theta_{f\,\textrm{hw};m}}(x) - F_{\textrm{all}_\textrm{hw};n}(x)
\right |
\geq \kappa_{\alpha} ,
\label{eq:cclSafe}
\end{equation}
with 
\begin{equation}
F_{\textrm{set;a}}(x) = 
\frac{1}{a} 
\sum_{i=1}^{a} 
\begin{cases}
1 & \textrm{if } x_i < x \\
0 & \textrm{if } x_i \geq x\\
\end{cases} ,
\end{equation}
where $x$ represents an SNR threshold,
$F_{\theta_{f\,\textrm{hw};m}}(x)$ represents the set of all $n$ hardware
injection artifacts identified with $CCL \geq 1$ and 
$F_{\textrm{all}_\textrm{hw};n}(x)$ are all $m$ known hardware injection
artifacts.  In applying our KS test, for a given significance
level($\kappa_{\alpha}$) one would accept or reject the specified null
hypothesis \cite{kstest2}.  If the left hand side (LHS) of equation~
\ref{eq:cclSafe} 
exceeds the chosen $\kappa_{\alpha}$ then one should accept the null
hypothesis because the artifact identified by the CCL
function are being preferentially selected from set of all known
hardware injections.  Acceptance of the null hypothesis means that the
sensor channel is unsafe.  In the case that
the LHS of equation \ref{eq:cclSafe} doesn't exceed $\kappa_{\alpha}$, the
sensor channel used to construct the CCL function is safe.  This
preliminary study required the boundary between safe and unsafe to
be set by $\kappa_{\alpha} \leq 0.85$, which means that at the $85\%$
confidence level we are sure of the safety or lack of safety for a
tested sensor channel.  This method of safety testing determined which
sensor channels are safe and suitable for analysis with the CCL method. 

\section{Preliminary Results}
We will show how the Critical Coupling Likelihood (CCL) method 
responds to artifacts in LIGO data, while ignoring effects that can be
attributed to potential GW signals.  The data used in this analysis is
a set of sampled instrument times, which were pre-processed with
a TF transformation tool called the omega pipeline.
This tool was chosen primarily because this transformation used a
wavelet decomposition basis which worked well for resolving signals
with low frequency components \cite{explainOmega}.  The total aggregate
time of our data samples is equivalent of 3.5
days of observing  data, derived from sampling two months near the
end of S6 for LIGO Livingston and LIGO Hanford observatories.

The costs of using CCL in the future are dependent on many factors.
The most costly phase of a
CCL end to end analysis is the data pre-processing step.  In order to
apply CCL to an online detector one would need at least 1 compute core
per every 5 to 10 sensor channels being analyzed.  This estimate is
reasonable if one assumes getting the sensor 
data to the compute core is instantaneous.  The actual computational
costs of the CCL method could be much higher depending on implementation of CCL,
the pre-processor and data handling at each
detector site.

The results presented here are intended to motivate the development of
the CCL method, highlight typical CCL functions, and show encouraging
results derived from a limited set of input sensors.  The reader
should bear in mind that these results are from a single case
study using CCL. There still remains a great deal of optimization, use of 
higher dimensional approaches (use of more artifact properties) and studies
with larger sets and fine tuning also can be performed.

The goal of a CCL analysis is to make quantitative
statements about the instrument behavior relative
to the studied sensor data.  The CCL output is a single quantitative
value, per available sensor data set, which is useful to determine the
coupling state of the detector.   The CCL functions have behaved
differently when applied to LIGO Livingston in comparison with LIGO 
Hanford.  This difference in behavior is unremarkable and likely
attributable to the distinctly different operating environments.

\begin{figure}
\centering
\subfloat[CCL Function LLO (Singlet)]
{\label{fig:LLO_Singlet}\includefigure[1.0\myfigwidth][!]
{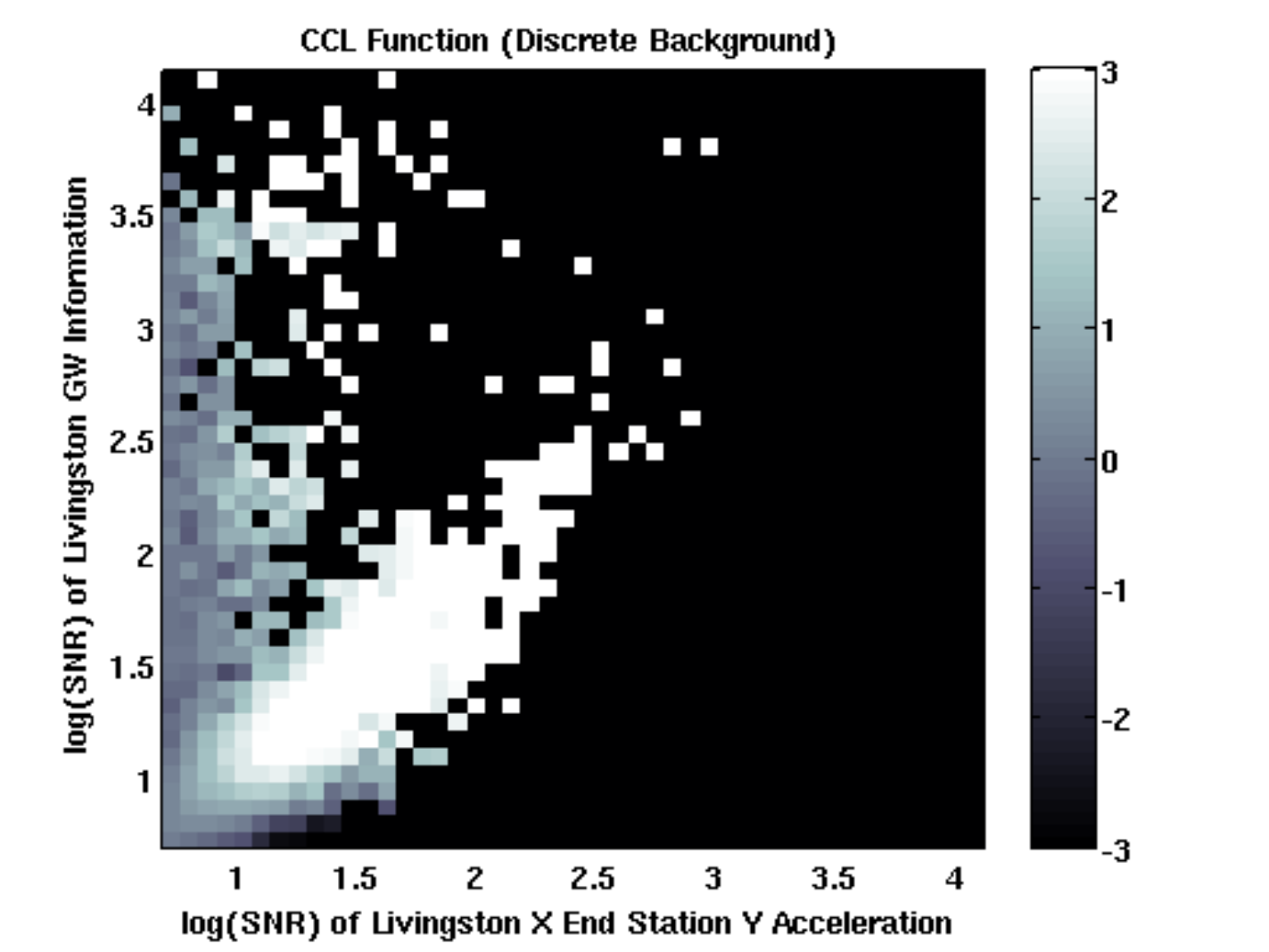}}
\\[0pt]
\subfloat[CCL Function Contour includes DQ flags]
{\label{fig:LLO_SingletDQ}\includefigure[1.0\myfigwidth][!]
{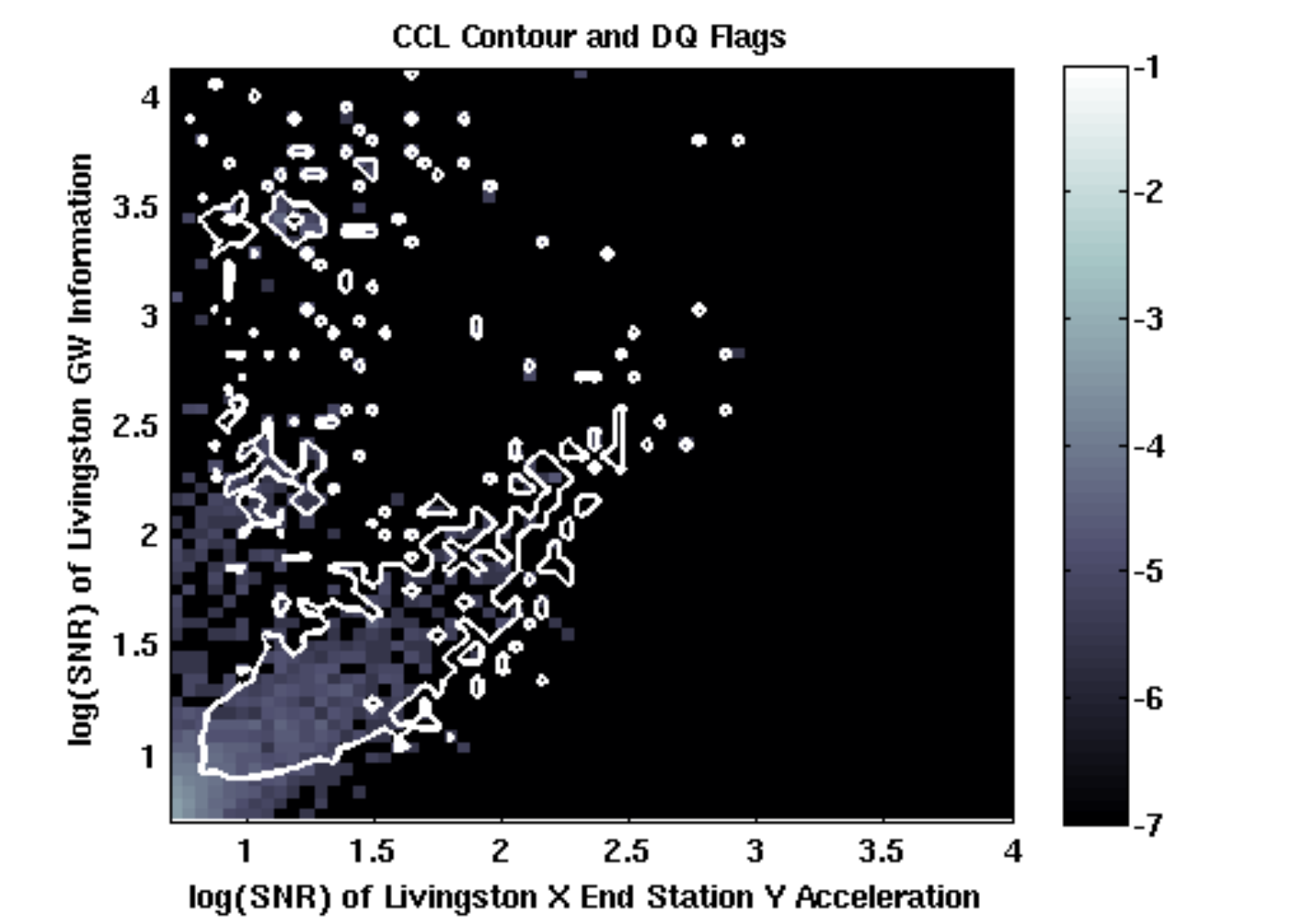}}
\caption{In \ref{fig:LLO_Singlet} we have an example of a ``Singlet''
  structure, which shows a log-linear correlation between artifacts in the
  sensor and GW artifacts.  In \ref{fig:LLO_SingletDQ} the same
  information is show as a shaded plot overlaid with contours denoting
  artifacts identified using current data quality analysis
  techniques\cite{ChannelSafety02}. One can notice that there is
  significant agreement between them.  The noise in this example is
  related to increasing pitch motion of a large optic at the end of
  one detector arm.  The accelerometer responds to the increasing 
  seismic motion because it is mounted on the assembly
  carrying the large optic. }  
\end{figure}

As an example LIGO Livingston 
is plotted in figure \ref{fig:LLO_Singlet}.  The identifiable
structure in this plot is an example of a \textit{Singlet}. The
singlet in this function is indicative of coupling to the instrument
through seismic motion. 
The seismic motion induces an acceleration of an accelerometer mounted
on an optics table in close proximity to one of LIGO's large
optics(test mass). This optic is pitching back and forth while
suspended at the end of one of LIGO's two arms as a result of optical
table motion.  The optics increasing motion corresponds with
increasing SNR of artifacts in the GW detector output.  This
relationship is readily seen in the figure as an angled structure of
high CCL values.   

\begin{figure}
\centering
\subfloat[CCL Function (Couplet)]
{\label{fig:LHO_Couplet01}\includefigure[1.0\myfigwidth][!]
{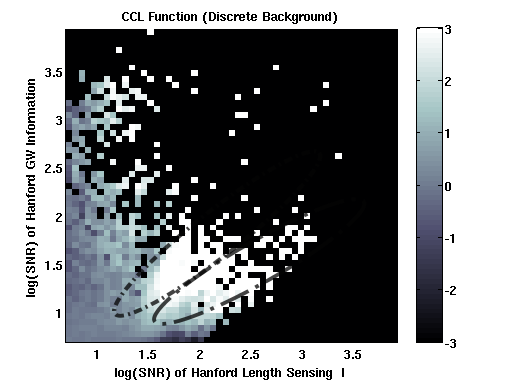}}
\\[0pt]
\subfloat[CCL Function Contour includes DQ
flags]{\label{fig:LHO_Couplet01DQ}\includefigure[1.0\myfigwidth][!]
{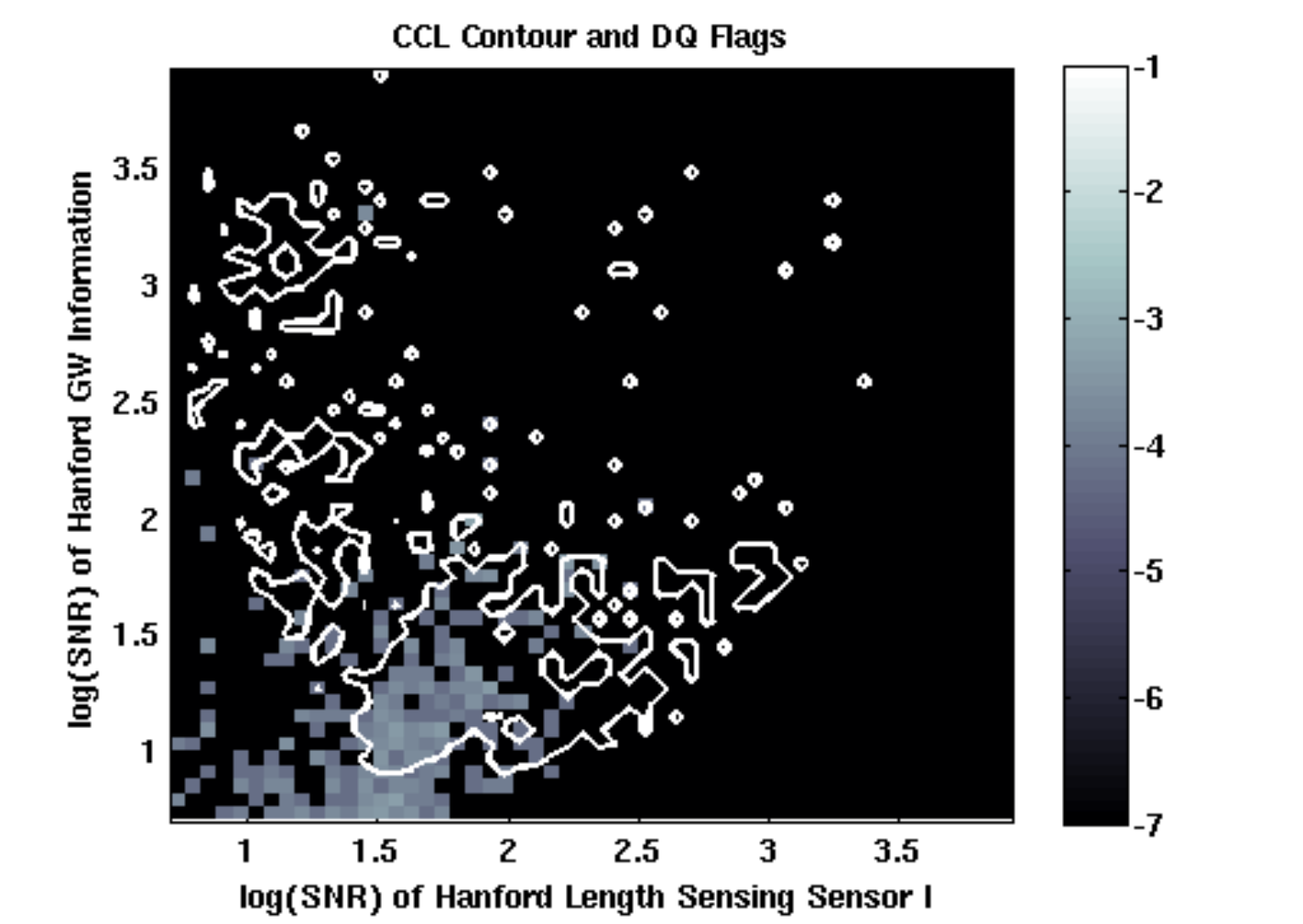}}
\caption{In \ref{fig:LHO_Couplet01} we have an example of a ``Couplet''
  structure, which shows bi-modal correlation between artifacts in the
  sensor and GW artifacts.  In
  \ref{fig:LHO_Couplet01DQ} the same CCL information is shown as a shaded
  plot overlaid with contours denoting artifacts identified using
  current data quality analysis techniques\cite{ChannelSafety02}.
  In the contour plot the standard techniques show agreement for one
  half of the bi-modal region.The other half is identified in figure
  \ref{fig:LHO_Couplet02DQ}. The noise shown for Hanford Length Sensor
  I(in-phase) is produced by 
  excitations of common motion of the input optics making up the two
  Fabry-Perot arms of the interferometer.} 
\end{figure}

\begin{figure}
\centering
\subfloat[CCL Function (Couplet)]
{\label{fig:LHO_Couplet02}\includefigure[1.0\myfigwidth][!]
{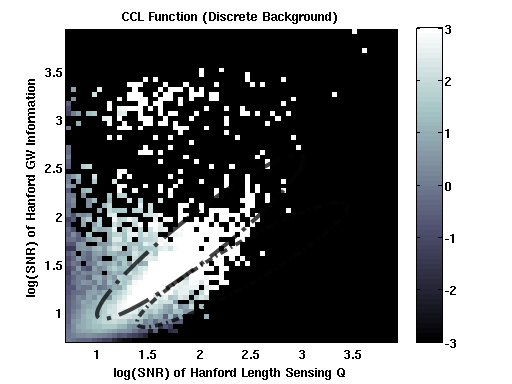}}
\\[0pt]
\subfloat[CCL Function Contour includes DQ flags]
{\label{fig:LHO_Couplet02DQ}\includefigure[1.0\myfigwidth][!]
{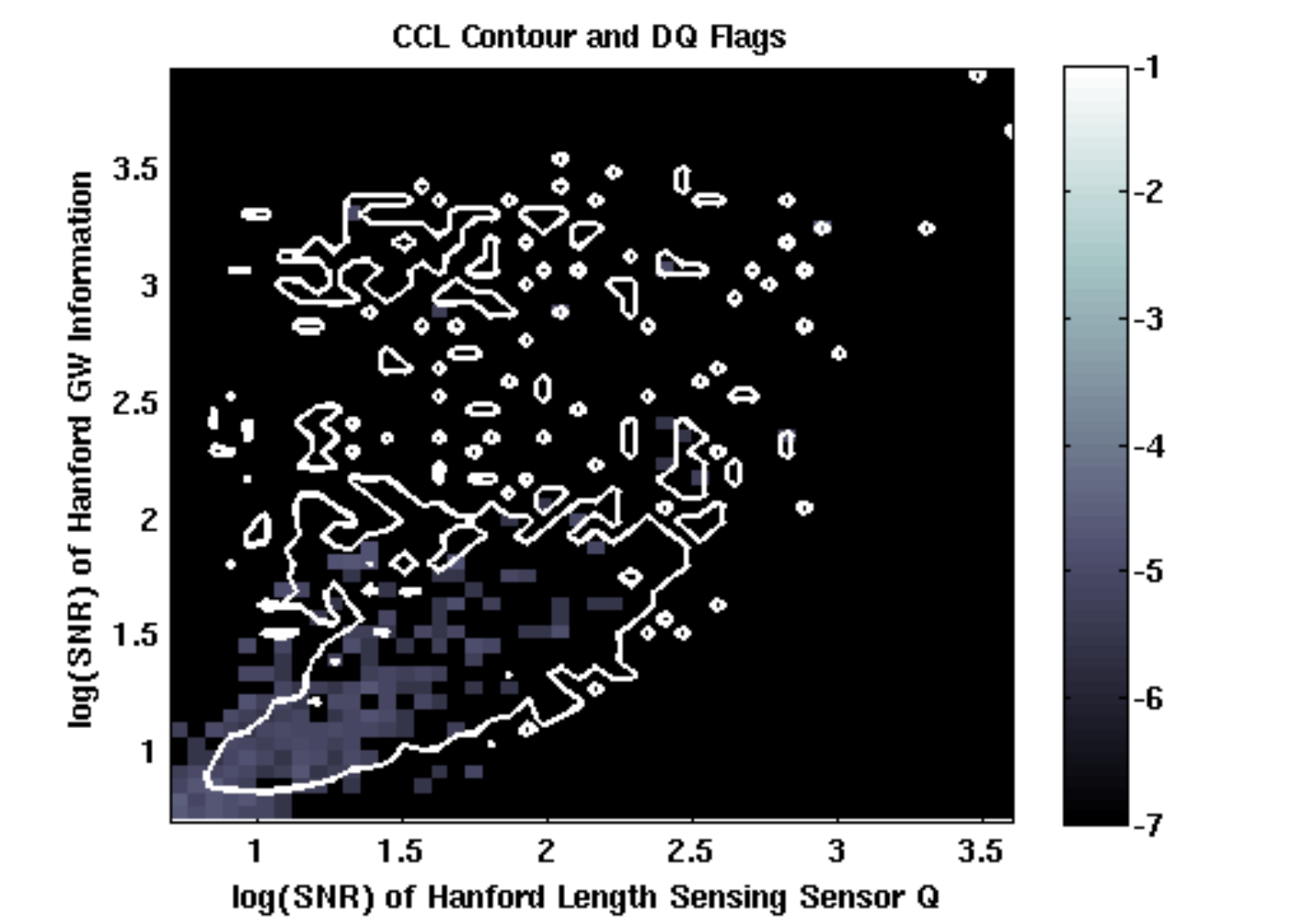}}
\caption{In \ref{fig:LHO_Couplet02} we have ``Couplet''
  structure shown in figure \ref{fig:LHO_Couplet01}.  In
  \ref{fig:LHO_Couplet02DQ} the CCL information is shown as a contour
  plot overlaid with artifacts identified using current data quality
  analysis techniques\cite{ChannelSafety02}.  The contour plot shows  
  agreement for one half of the bi-modal region.  The other half is
  identified in figure \ref{fig:LHO_Couplet01DQ}. The noise shown
  using Hanford Length Sensor Q(quadrature) is produced by excitations
  of a differential motion between the input optics that form the
  Fabry-Perot arms of the interferometer.} 
\end{figure}

Another type of CCL structure example is shown using LIGO Hanford
data in figures \ref{fig:LHO_Couplet01} and \ref{fig:LHO_Couplet02}.
These figures shows a different type of identifiable structure, a
``Couplet''.  For CCL functions with this type of structure, the
artifacts identified by this model can typically be associated with
another sensor. In both figures, there appears to be two angled
structures, not always easily distinguishable, but in this case one
structure is better defined than the other.  In figure
\ref{fig:LHO_Couplet01}, the well defined angled structure(short
dashes) corresponds to a poorly defined, but similar 
structure in figure \ref{fig:LHO_Couplet02}.  This relationship is
reciprocal, the well defined angled structure(long dashes) associated
with figure \ref{fig:LHO_Couplet02} is the poorly defined 
structure associated with the couplet from figure
\ref{fig:LHO_Couplet01}.  The couplet structure appears in pairs of
CCL functions.  These pairs are usually related by obvious mechanisms
as in demodulated signals or signals derived from mechanically
interdependent systems.  For these example figures, the two data 
channels are registering artifacts which have both
in-phase and quadrature components of light received by the same
photo-diode assembly.  The in-phase component of light tracks common
motion of the input optics which form the long Fabry-Perot arms of the
interferometer.  The quadrature phase tracks the differential motion
of the input optic of the cavity.  Laser power fluctuations can induce
optic motion impacting the length of the Fabry-Perot cavities.
These length changes can take the cavity off resonance and allow energy
to exit the cavity.  The energy exiting the cavity can transfer power
from one arm to another leading to differential motion of the input
optics.  These two types of motion, common and differential, can
excite one another and make controlling the full length interferometer
difficult thereby increasing the noise in the interferometer.

The CCL method can provide a quantified measure describing coupling into
the detector data.  This measure is useful to understand if and what 
local environmental effects may be responsible for a false GW event.
To appreciate the potential usefulness of a method 
like the CCL, we imposed a CCL threshold value.  Thus exceeding this
value declares an event as noise coupled to the detector.  By using a
CCL cutoff, $CCL \geq 1$, it allows us to sort the identified
artifacts into two groups: inherent noise artifacts (uncoupled) and
those which are due to environmental effects (coupled).  
This CCL value corresponds to an identified artifact is at least
$\sqrt{10}$ times more likely to be coupled than uncoupled, as defined
in section \ref{sec:calculationCCL}.

\begin{figure}
\centering
\includefigure[\textwidth]{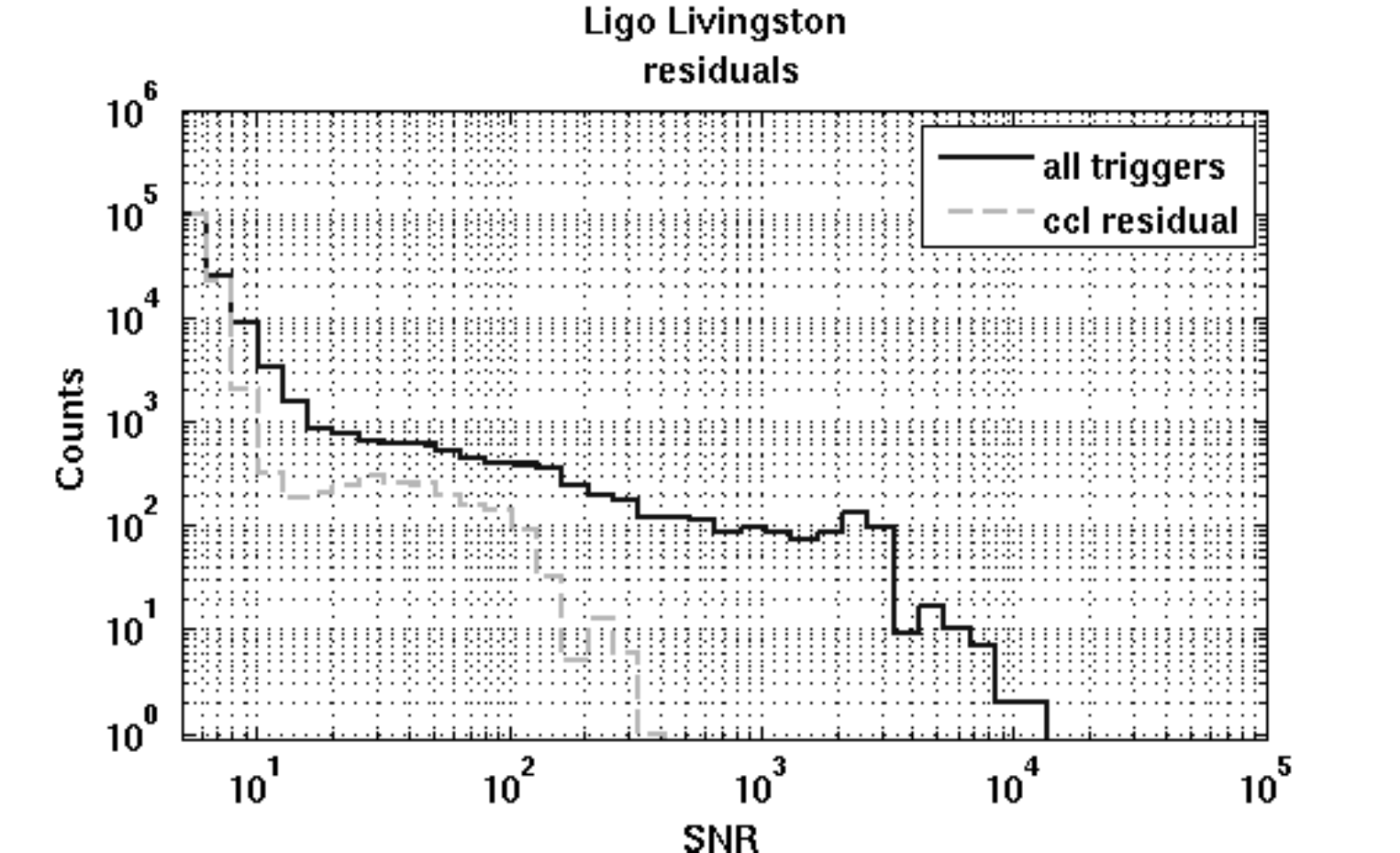}
\caption{This figure shows the distribution of LIGO Livingston GW
  artifacts, as a solid curve.  The dotted curve is set of remaining 
  artifacts from the sampled set (solid curve) after removing
  $(CCL \geq 1)$ identified artifacts.}
\label{fig:LLOresidual}
\end{figure}

\begin{figure}
\centering
\includefigure[\textwidth]{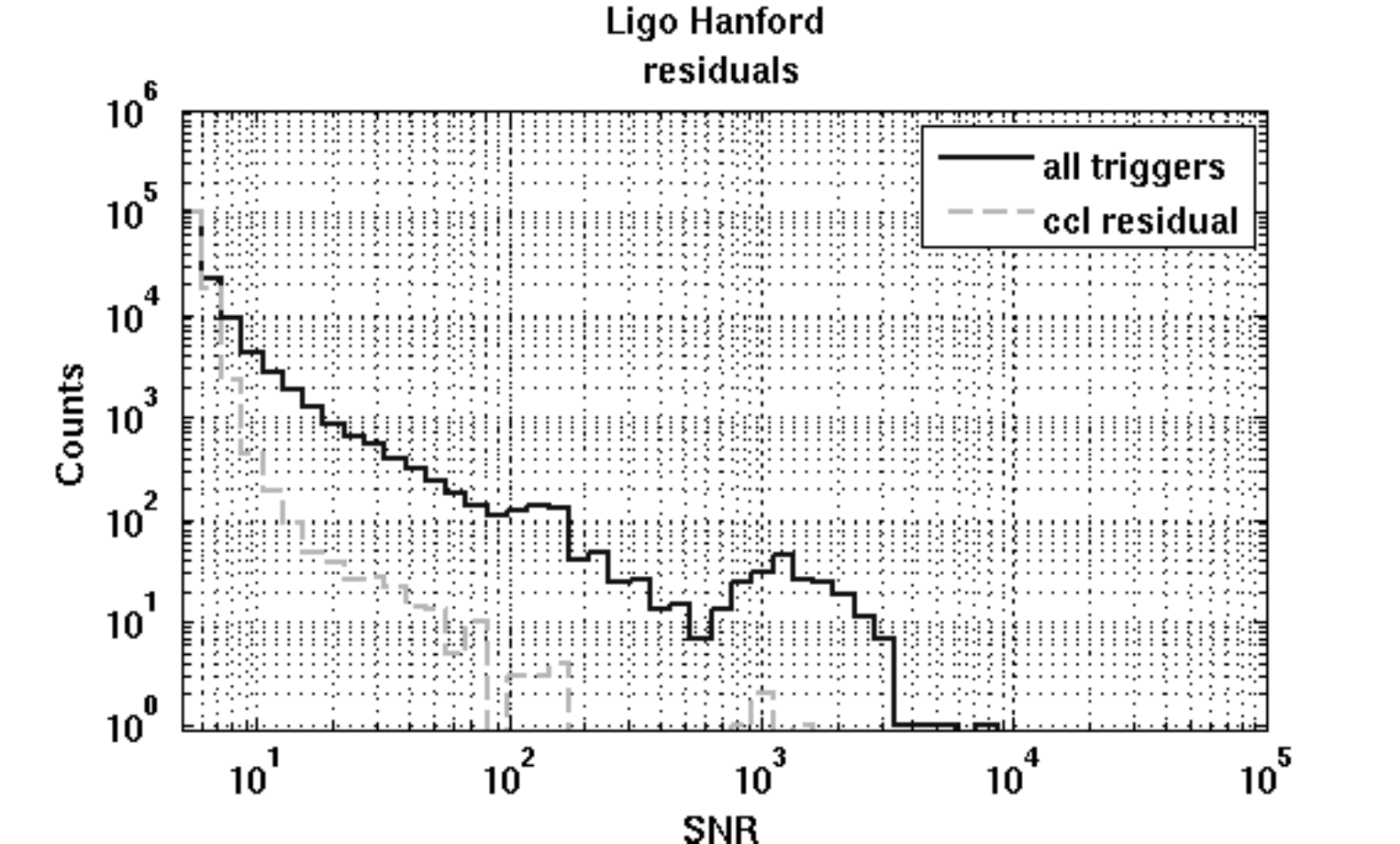}
\caption{This figure shows the distribution of LIGO Hanford GW
  artifacts, as a solid curve.  The dotted curve is set of remaining 
  artifacts from the sampled set (solid curve) after removing $(CCL
  \geq 1)$ identified artifacts.}
\label{fig:LHOresidual}
\end{figure}

For LIGO Livingston, a total of 205 functions were used to
identify the coupled artifacts.  For LIGO Hanford, we used a total 
of 207 functions to identify the coupled artifacts in the
Hanford data set.   The set of artifacts analyzed were 
constructed by sampling Livingston and Hanford data with an average
sampling rate of 1 sample, which is 1 second in duration, every two
minutes using only science mode times.  Science mode times are periods
where the instrument is operating at the minimum level of data quality
that should be considered viable to perform a search for GW
signals \cite{HowToDQ}.  Using this sampling constraint, the data set is less 
than $5\%$ of the total amount of data available from two months near
the end of LIGO's sixth science run.  We expect that if more samples were
taken during the same interval of observation time the
results of our preliminary analysis will remain relatively unchanged.
Using these constraints on sampling rate and sensors used, figures
\ref{fig:LLOresidual} and \ref{fig:LHOresidual} show the number of
artifacts not identified as coupled after applying a CCL cut.  For
both figures, there is a 
clear suppression of the original outlier tails.  What may not be
readily apparent in these figures is that the CCL method appears able to
suppress the outlier tail and identify moderate SNR artifacts, while
leaving the original low SNR artifact distribution unchanged.  In table
\ref{tab:suppression}, it is easier to see why one can appreciate the
CCL method for identifying artifacts, because this method ignores SNR
regions where artifacts tend to be the result of inherent noise properties
of the instrument.  This suppression of the outlier tail in this study
is accompanied by a data volume cost.  The amount of gravitational
wave data that should be discarded, at least for this single case,
is moderately high at $13.9\%$ and $15.1\%$ of the tested data for LIGO
Livingston and LIGO Hanford respectively.  A LIGO type detector has
noise sources which are clearly non-Gaussian in nature.  Some of these 
excesses have been identified, and are more pronounced in 
LIGO's low frequency bands of $(0\textrm{Hz},
1\textrm{kHz}]$, which is thought to account for on order of $10\%$,
of the excess outlier tail \cite{DataQuality}.  The artifacts
contributing to the discarded data show a clear
selection of non-Gaussian type artifacts.  It is likely that our 
observed discarded data volume is not the result of our analysis
configuration, but may actually be the product of an epoch which is
less well behaved than one might expect.  Further studies using CCL
are called for but with this single case study there is already
evidence that this method may prove to be beneficial for GW
searches. 

\begin{table}
\begin{minipage}[c]{1.0\textwidth}
\centering
\begin{tabular}{|c|c|c|}
\hline
 & Livingston & Hanford \\
\hline
Deadtime & 13.9\% & 15.1\% \\
\hline \hline
SNR($\geq$) & \% & \% \\
\hline
5 & 14.4 & 16.3 \\
8 & 82.3 & 93.5 \\
10 & 88.8 & 96.7 \\
20 & 86.7 & 98.0 \\
50 & 92.4 & 98.9 \\
100 & 98.0 & 99.2 \\
\hline
\end{tabular}
\caption{This is the efficiency and deadtime table for the Livingston
  and Hanford observatories.  The deadtime estimated presented here is
  expected to be an upper limit on the expected deadtime for
  $(\textbf{CCL} \geq 1)$.  Deadtime is analysis time which should
be discarded because it is suspected of having to poor data quality
characteristics \cite{HowToDQ}.} 
\label{tab:suppression}
\end{minipage}
\end{table}

The sensor channels used in this preliminary study were those of the
standard follow-up activities done at the end of a GW 
search \cite{FollowupWorkshop,flwupGouaty}.  Using the safety
prescription outlined 
in section \ref{sec:disregardCCL} we discarded several channels
presenting undesirable sensitivity to GW phenomena and considered them
potentially unsafe.  According to our chosen criteria, we were
required to remove 10  channels for LIGO Hanford, and 9 channels for
LIGO Livingston.  Of the channels removed, a large fraction were 
consistent with channels suspected to be unsafe, with a
few exceptions.  One example unsafe channel noted for both
detectors was the channel which records the amount of light incident
on the detector's output mode cleaner.  The light that is 
incident on the mode cleaner should contain the signature of a passing GW
so, it would be unwise to use this channel as an indicator of detector
noise.  Not all channels failing to meet the safety criteria were
common to both detectors. In general the types of channels
that were consistently seen as unsafe, were either related to the
output of the detector, or part of the feedback control of the
detector.

One can see in table \ref{tab:safeChannels}, that the remaining safe
channels identify a small fraction of the total number 
of hardware injection artifacts.  There is a notable contrast between
the safe and unsafe functions, because as one can see from
tables \ref{tab:safeChannels} and \ref{tab:unsafeChannels} 
the median number of hardware injections identified per CCL function
for safe channels is lower than those for unsafe channels.

\begin{table}
\centering
\begin{tabular}{|r|c|c|}
\hline \hline
\multicolumn{3}{|c|}{\textbf{Safe} CCL Function Properties} \\
\hline
  & Livingston & Hanford \\
\hline \hline
Median Injection Count Per Channel & 2 & 4\\
\hline
Mean Injection Count Per Channel & 2.7(1.04\%) & 4.7(1.54\%) \\
\hline
Standard Deviation on Count Per Channel & 2.3(0.87\%) & 4.2(1.39\%) \\
\hline
Hardware injection artifacts in data set & 265 & 303 \\
\hline
Total Number of Channels & 205 & 207 \\
\hline
\end{tabular}
\caption{This table shows the average number of hardware injection artifacts
  identified per safe.  The hardware injection artifacts identified by
  safe channels were found not to be consistent with population of all
  hardware injections performed.}
\label{tab:safeChannels}
\end{table}

\begin{table}
\centering
\begin{tabular}{|r|c|c|}
\hline \hline
\multicolumn{3}{|c|}{\textbf{Unsafe} CCL Function Properties} \\
\hline
  & Livingston & Hanford \\
\hline \hline
Median Injection Count Per Channel & 7 & 9\\
\hline
Mean Injection Count Per Channel & 6.8(2.58\%) & 9.2(3.04\%) \\
\hline
Standard Deviation on Count Per Channel & 1.2(0.47\%) & 3.1(1.05\%) \\
\hline
Hardware injection artifacts in data set & 265 & 303 \\
\hline
Total Number of Channels & 9 & 10 \\
\hline
\textbf{Ph.D. Thesis, MIT Dept. of Physics}
\end{tabular}
\caption{This table shows the average hardware injection artifacts
  identified per unsafe channel.  The hardware injections identified
  are consistent with all hardware injections performed implying
  that these channels are1 unsafe for use to determine noise coupling
  in the detector. }  
\label{tab:unsafeChannels}
\end{table}

\section{Conclusions and Future Directions}
The Critical Coupling Likelihood method is intended to improve future
GW search efficiency and provide feedback to
instrument scientists at the observatory.  The results shown here 
are preliminary and are produced using a small set of LIGO data.
The results shown use only the simplest form of CCL functions,
capitalizing on the properties SNR and coincidence.  We
expect to see this same level of performance if we apply this
approach to a full set of existing LIGO data or future 
aLIGO data.  This method is promising because it offers
quantified detector information which can be directly imported into
a GW search.  In creating this quantified detector information, the
CCL method is also providing instrument scientists with 
information that describes potential coupling mechanisms between the
instruments operating environment and behavior.  We
plan to present the instrumental implications of the CCL method in a
later paper.

In this paper we introduced a new detector characterization technique
called the CCL method.  This method is
intended to use all available observatory information and deduce the
presence of coupling between the environment and the detector.  In
making this coupling 
identification it also implicitly estimates the coupling relationship
giving information about how the strength of local environmental effects
map into specific noise levels and frequency bands.  These preliminary
results indicate that an approach like CCL
should have an appreciable impact on the effectiveness of future GW
searches, by reducing the outlier tail. Our initial results show CCL
identifying $\sim 80\%$  of observed artifacts with $SNR \geq 8$.
Identifying these outlier events leads to significant outlier tail
suppression which is clearly seen in figures \ref{fig:LLOresidual} and
\ref{fig:LHOresidual}. By using this technique we can quantitatively
characterize individual GW candidates, accounting for instrumental
conditions, and hence improve GW searches and the validation of a
detection.

\section{Acknowledgments}
The authors gratefully acknowledge the support of the United States
National Science Foundation for the construction and operation of the
LIGO Laboratory.  The authors also thank  LIGO and the LIGO Scientific
Collaboration for allowing the use of LIGO data for this study.  The
authors would also like to acknowledge the support of this research
via NSF grants PHY-0905184 and PHY-0757058.  C Costa also thanks to
CNPq 300001/2012-6 by financial support. This paper was assigned
LIGO document number LIGO-P1000187.

\end{document}